\documentstyle[psfig]{mn}
\newif\ifAMStwofonts                        
\def\lsimeq{{_<\atop^{\sim}}}
\def\gsimeq{{_>\atop^{\sim}}}

\def\eso36{\hbox{ESO~3.6-m}}                          
\def\nin{\noindent}

\title[Radio-MIR Correlation]{The Radio--Mid-Infrared Correlation and the
Contribution of 15-$\mu m$ Galaxies to the 1.4-GHz Source Counts}
\author[C. Gruppioni, F. Pozzi, G. Zamorani et al.]
{\parbox[]{6.5in} {C. Gruppioni$^{1,2}$\thanks{e-mail: gruppioni@bo.astro.it}, 
F. Pozzi$^{2,3}$, G. Zamorani$^{2,4}$, P. Ciliegi$^{2}$, C. Lari$^{4}$, 
E. Calabrese$^{2}$, F. La Franca$^{5}$ and I. Matute$^{5}$} \\ \\
$^{1}$ Istituto Nazionale di Astrofisica: Osservatorio Astronomico di Padova, vicolo dell'Osservatorio 5, I--35122 Padova, Italy \\
$^{2}$ Istituto Nazionale di Astrofisica: Osservatorio Astronomico di Bologna, via Ranzani 1, I--40127 Bologna, Italy\\
$^{3}$ Dipartimento di Astronomia, Universit\`a di Bologna, via Ranzani 1, I--40127 Bologna, Italy\\
$^{4}$ Istituto di Radioastronomia del CNR, via Gobetti 101, I--40129 Bologna, Italy\\
$^{5}$ Dipartimento di Fisica, Universit\`a ``Roma Tre'', via della Vasca Navale 84, I--00146 Roma, Italy\\
}
                   
\date{Accepted 2003 March 4. Received -----; in original form -----.}

\pagerange{\pageref{firstpage}--\pageref{lastpage}}

\pubyear{2002}

\def\LaTeX{L\kern-.36em\raise.3ex\hbox{a}\kern-.15em
  T\kern-.1667em\lower.7ex\hbox{E}\kern-.125emX}

\begin{document}

\small
\label{firstpage}

\maketitle                  
      
\begin{abstract}
  The radio counterparts to the 15-$\mu$m sources in the European Large
  Area ISO Survey southern fields are identified in 1.4-GHz maps down
  to $\sim 80~ \mu$Jy. The radio - MIR correlation is investigated and
  derived for the first time at these flux densities for a sample of
  this size. Our results show that radio and MIR luminosities
  correlate almost as well as radio and FIR, at least up to $z \simeq
  0.6$. Using the derived relation and its spread together with the
  observed 15-$\mu$m counts, we have estimated the expected
  contribution of the 15-$\mu$m extragalactic populations to the radio
  source counts and the role of MIR starburst galaxies in the well
  known 1.4-GHz source excess observed at sub-mJy levels. Our analysis
  demonstrates that IR emitting starburst galaxies do not contribute
  significantly to the 1.4-GHz counts for strong sources, but start to
  become a significant fraction of the radio source population at flux 
  densities $\lsimeq 0.5 - 0.8$ mJy. They are expected to be responsible for
  more than 60\% of the observed radio counts at $\lsimeq 0.05$ mJy.
  These results are in agreement with the existing results on optical
  identifications of faint radio sources.
\end{abstract}


\section{Introduction}
With a thousand times better sensitivity than the $IRAS$ 12-$\mu$m
data, the LW3 observations (in the 12 -- 18 $\mu$m waveband, centred at
$\lambda =$15 $\mu$m) with the $ISOCAM$ camera (Cesarsky et al. 1996)
on the {\it Infrared Space Observatory} ($ISO$, Kessler et al. 1996)
have allowed us for the first time to perform sensitive surveys of
distant infrared galaxies (up to $z \sim 1.5$) in the Mid-Infrared
(MIR) band. The extragalactic source counts derived from these
15-$\mu$m surveys, including large area shallow surveys like ELAIS
(Oliver et al. 2000) covering the flux density range $0.5 \leq S_{15~\mu m}
\leq 150$ mJy (Lari et al. 2001), and small area deep integrations
reaching $S_{15~\mu m} \simeq 0.05 - 0.1$ mJy (Elbaz et al. 1999), show
a strong departure from no evolution predictions at low flux densities
($\leq 1-2$ mJy; Elbaz et al. 1999; Gruppioni et al. 2002).  According
to both optical identification works (Aussel et al. 1999; Elbaz et al.
1999; Pozzi et al. 2003) and theoretical models (i.e.
Franceschini et al. 2001), the sources responsible for the sharp upturn
observed in the number counts at faint flux densities are mainly star-forming
galaxies at moderately high redshifts (0.4$ \leq z \leq$ 1.4).

A tight correlation between Far-Infrared (FIR) and radio continuum for
star-forming galaxies is locally well assessed over a large range of
luminosities, from normal spirals to the more extreme Ultraluminous
Infrared Galaxies (ULIGS: $L > 10^{12} L_{\odot}$), as shown by several
authors (Condon 1992; Cram et al. 1998; Yun, Reddy \& Condon 2001). So
far there have been many attempts to explain the tightness of this
correlation, whose origin is still somewhat unclear. It is generally
assumed that massive stars are responsible for both the UV photons
heating the dust, which re-radiates in the infrared band, and the
acceleration of relativistic electrons, producing the radio continuum,
after their explosion as supernovae. Such a tight local correlation
does not necessarily hold in the distant Universe and, even in this
case, it would be interesting to investigate possible variations with
redshift of the slope and/or normalization of the correlation.
Moreover, it is not obvious that the radio continuum should correlate
with the MIR emission (produced by a mixture of stochastic heating from
Polycyclic Aromatic Hydrocarbons (PAH; i.e. Puget \& Leger 1989) and thermal 
emission at high
temperature) as well as with the FIR (resulting from thermal emission
of large dust grains at lower temperature).

The aim of this paper is to study for the first time the radio--MIR
correlation for galaxies at cosmological redshifts through a
statistically significant sample. The considered sample contains 65 sources 
detected at both 15 $\mu$m and 1.4 GHz in the ELAIS southern fields, 
all identified with galaxies up to $z \sim$0.8. By introducing the radio--MIR
correlation into the evolutionary model fitting the 15-$\mu$m
extragalactic source counts (see Gruppioni et al. 2002), we have then
estimated the expected contribution to the radio source counts from the
MIR star-forming galaxies, whose role and importance relative to that
of AGN in the 1.4-GHz excess of sub-mJy/$\mu$Jy radio sources is still
matter of debate.

\begin{figure*}
\begin{minipage}{17.5cm}
\vspace{-0.2cm}
\centerline{\psfig{figure=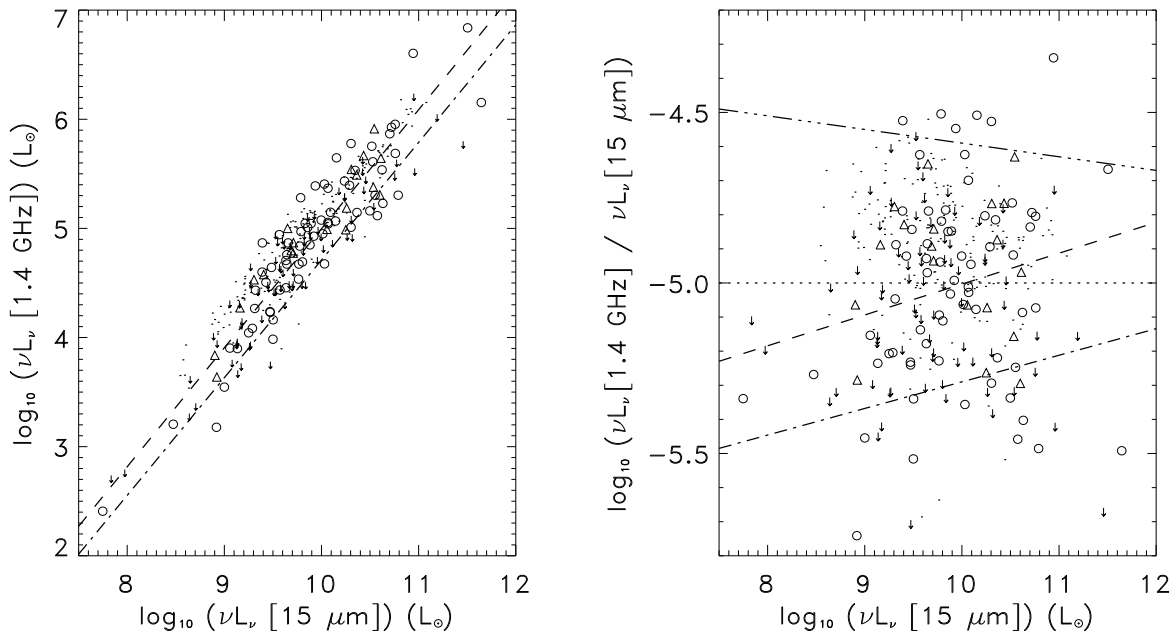,width=17cm,angle=90}}
\caption{$Left$ -- 
Radio (1.4 GHz) versus MIR (15 $\mu$m) luminosity for the 15-$\mu$m 
galaxies in the ELAIS southern fields. Objects detected in the radio
are represented by empty circles (galaxies with spectroscopic redshift) and
triangles (without spectroscopic information, to which a redshift has been
assigned on the basis of an empirical relation between MIR flux density and 
$z$: see text).
Arrows are the 3$\sigma$ radio upper limits with measured redshift, while 
the small dots are the 3$\sigma$ radio upper limits with no spectroscopic $z$.
The dashed line is the best-fitting relation obtained considering only the
radio detections with spectroscopic redshift, while the dot-dashed line is the best-fitting
relation obtained by taking into account also the radio upper limits (and objects without
spectroscopic redshift) through the ASURV package (Isobe, Feigelson \& Nelson 1986).
$Right$ -- Radio to MIR luminosity ratio versus MIR luminosity. Symbols, as well as the dashed
and dot-dashed lines, are the same as in the $left$ panel. The dotted line is the local 
radio--FIR correlation converted into radio--MIR through the empirical relations between 
$L_{FIR}$ and $L_{15~\mu m}$ found locally by Elbaz et al. (2002). The 
dot-dot-dot-dashed line is the best-fitting relation found by Garrett et al. (2002)
for a radio--MIR sample in the HDF-N.
} 
\vspace{-0.3cm}
 \label{fig_ricorr}
\end{minipage}
\end{figure*}

The paper is structured as follows. In section \ref{data} we present
our data sample. In section \ref{radio_mir_corr} we derive the
radio--MIR correlation and investigate its redshift dependence. In
section \ref{radio_cnt} we estimate the contribution of infrared
star-forming galaxies to the radio source counts and discuss the
results and implications. In section \ref{concl} we present our
conclusions.

Throughout this paper we will assume $H_0 = 75$ km s$^{-1}$ Mpc$^{-1}$, $\Omega_m = 0.3$
and $\Omega_{\Lambda} = 0.7$.

\section{The Data Samples}
\label{data}
The ELAIS survey at 15 $\mu$m (Oliver et al. 2000), performed in raster
mode with the ISOCAM instrument, covers a total area of $\sim$ 12
deg$^{2}$ divided into 4 main fields and several smaller areas. One of
the main fields, S1, and one of the smaller areas, S2, are located in
the southern hemisphere. S1 is centred at $\alpha$(2000) = 00$^h$
34$^m$ 44.4$^s$, $\delta$(2000) = $-43^{\circ}$ 28$^{\prime}$
12$^{\prime \prime}$ and covers an area of $2^{\circ} \times
2^{\circ}$, while S2 is centred at $\alpha$(2000) = 05$^h$ 02$^m$
24.5$^s$, $\delta$(2000) = $-30^{\circ}$ 36$^{\prime}$ 00$^{\prime
  \prime}$ and covers an area of $21^{\prime} \times 21^{\prime}$.  The
15-$\mu$m data in these fields have been reduced and analysed using the
{\it LARI technique}, especially developed for the detection of faint
sources (Lari et al. 2001), obtaining two samples at $\geq 5\sigma$
containing 462 sources with $0.5 \lsimeq S_{15 \mu m} \lsimeq 150$ mJy
in S1 (available at {\sf
  http://www.bo.astro.it/$\sim$elais/catalogues/ELAIS\_CAM\_15micron\_\\\_S1.TAB})
and 43 sources with $S_{15 \mu m} \gsimeq 0.4$ mJy, in the deeper
field S2 respectively. The 15-$\mu$m source counts have been derived by
Gruppioni et al. (2002) using the $\sim$320 extragalactic sources
detected over the S1 area.

The whole S1 and S2 areas have been surveyed in the radio with the
Australia Telescope Compact Array to $S_{1.4 GHz} \simeq$ 0.2 and 0.13
mJy respectively (Gruppioni et al. 1999; Ciliegi et al. 2002, in
preparation) and in several optical bands: S1 in the $R$ band (to $R
\sim 22.5$) with the ESO/Danish 1.5-m Telescope and S2 in the $U$, $B$,
$R$ and $I$ bands with the WFI at the ESO 2.2-m Telescope and in the
$K^{'}$ band with SOFI at the ESO NTT.
The radio catalogue in S1 (available at {\sf
  http://www.bo.astro.it/$\sim$elais/catalogues/ELAIS\_RADIO\_S1.TAB})
consists of 652 1.4-GHz sources detected at the $5\sigma$ level,
while the S2 radio catalogue consists of 75 sources.

Spectroscopic observations of the optical counterparts of the ISOCAM
sources were carried out at the 2dF/AAT and ESO Danish 1.5-m, 3.6-m and
NTT telescopes. In S1 all the sources with R$<$21.0 have been
spectroscopically identified ($\sim$210 extragalactic objects; La
Franca et al. 2002, in preparation). In S2 we have spectroscopic
informations for 29 sources brighter than $R \simeq 21$ (about 68\% of
the sample; Pozzi et al 2003) obtained with the ESO
3.6-m telescope.

\section{The Radio-MIR Correlation}
In order to investigate the radio--MIR correlation within our ISOCAM
sample, we have first cross-correlated the 15-$\mu$m and the 1.4-GHz
catalogues (complete at the 5$\sigma$ level) in S1 and S2, finding 28
and 13 coincidences respectively within a distance of 5 arcsec. Note
that the S1 15-$\mu$m catalogue considered for this analysis is the
same, conservative, catalogue used to derive the source counts (see
Gruppioni et al. 2002), where 35 possibly spurious sources (all with $S
< 1.5$ mJy) have been excluded after a visual inspection of their pixel
history. These sources, detected above the 5$\sigma$
threshold on the maps obtained through a combination of several images, 
are too faint to be distinguished from noise on the single pixel histories 
without uncertainty (see Gruppioni et al. 2002 for further explanation).
Then, at each ISOCAM position, we have searched for
detection in the radio maps down to 3 $\sigma$, finding 53 additional
radio identifications within 5 arcsec (46 in S1 and 7 in S2), for a
total of 94 ISO--radio associations with $S_{1.4~GHz} \geq 3 \sigma$.
The number of expected spurious coincidences, on the basis of the
density of ISO and radio sources and of the adopted maximum distance (5
arcsec) is of the order of one. Seventy-three of these sources have
spectroscopic data and redshift measurements in the spectroscopic data
sample available by October 2002 (additional spectra taken in the
October run were not used in this work).

In Table \ref{radio_det_tab} we present the percentage of radio
detections as a function of the 15-$\mu$m flux density. The first two
columns give the flux density range and the corresponding average flux 
density respectively, whilst the following columns give the total number of
ISOCAM extragalactic objects, the number of radio detections, 
the number of non-AGN radio detections(in brackets) and the corresponding
fraction (i.e. radio detections over total number of extragalactic
sources in that bin).
\label{radio_mir_corr}
\begin{table} 
\centering
  \caption{15-$\mu$m sources detected at 1.4 GHz}
  \label{radio_det_tab}
\begin{tabular}{crrrr}
 \hline
  $S$ (mJy) & $\left< S \right>$  & $N$ &   $N_{det}$   &  \%   \\
 \hline
 $< 0.5$    &   $-$               &  10       &   0  ~(0)     &   0.0 \\
 $0.5-0.9$  &   0.7               &  78       &  11  (11)     &  14.1 \\
 $0.9-1.6$  &   1.2               & 120       &  19  (17)     &  15.8 \\
 $1.6-2.9$  &   2.2               &  87       &  28  (27)     &  32.2 \\
 $2.9-5.3$  &   3.9               &  33       &  14  (11)     &  42.4 \\
 $5.3-9.5$  &   7.0               &  15       &  12  (11)     &  80.0 \\
$~9.5-17.0$ &  12.7               &   7       &   6  ~(5)     &  85.7 \\
$17.0-30.6$ &  22.8               &   2       &   2  ~(1)     & 100.0 \\
$30.6-55.1$ &  41.1               &   2       &   2  ~(1)     & 100.0 \\
 \hline
\end{tabular}
\end{table}
At high flux densities all the 15-$\mu$m extragalactic sources have a
radio counterpart, while the fraction of radio identifications
decreases at lower 15-$\mu$m flux densities. This is due to the
fact that our radio maps are not deep enough to allow the detection of
all our 15-$\mu$m sources, especially the fainter ones.
Figure \ref{fig_ricorr} ($left$ panel) shows the 1.4-GHz luminosity
versus the 15-$\mu$m luminosity for the 65 ISOCAM galaxies with
spectroscopic identification detected in the radio band (open circles).
As radio K-correction we have applied a power law with a slope
$\alpha$=0.7, whilst at 15 $\mu$m we have applied the K-corrections derived 
by Franceschini et al. (2001) using different template Spectral Energy 
Distributions (SEDs) for the different populations (M82 for starbursts and 
type 2 AGN, M51 for normal spirals) modelling the 15-$\mu$m source counts.

A formal fit to the observed radio--MIR luminosity
correlation for the radio detected galaxies only (dashed line) yields
\begin{eqnarray}
 ~~log (L_{1.4~GHz}/L_{\odot}) = (1.09 \pm 0.05) log (L_{15~\mu m}/L_{\odot}) \\\nonumber
  - (5.91 \pm 0.54)
\end{eqnarray}
\nin with a dispersion of $\sim$0.27 dex. Slope, normalization and
dispersion are consistent, within errors, with those of the local
determination of the radio--FIR relation found for $IRAS$ galaxies by
Yun, Reddy \& Condon (2001; $log (L_{1.4~GHz}) \propto (0.99 \pm 0.01)
log (L_{60~\mu m}); ~\sigma=0.26$). Since our determination does not
take into account the radio upper limits (about 3/4 of the total), it
is more representative of the upper envelope (i.e. the brighter radio
objects) than of the ``real'' radio--MIR luminosity distribution. In
order to include the MIR galaxies not detected in the radio, we
have recomputed the radio--MIR luminosity correlation for the entire
sample of 331 MIR-selected galaxies by using ASURV (the Survival
Analysis Package which uses the routines described in Isobe, Feigelson
\& Nelson 1986 and takes into account also the upper or lower limits in
a sample). A redshift has been assigned to the 192 sources without
spectroscopic information by using the empirical correlation (and its
spread) between the 15-$\mu$m flux densities and redshifts found for our 
139 spectroscopically identified galaxies:
\begin{equation}
log (z) = - (0.68 \pm 0.04) - (0.37 \pm 0.06) log (S_{15~\mu m} [mJy]) + G(0,\sigma_{rel})
\label{z_mir_corr}
\end{equation}
\noindent where $\sigma_{rel}$ (= 0.25) is the 1$\sigma$ dispersion of the relation and
$G(0,\sigma_{rel})$ is a Gaussian distribution with centre 0 and width $\sigma_{rel}$.
For the extragalactic sources not detected in the radio we have adopted an upper limit to
the 1.4-GHz flux density equal 
to their corresponding 3$\sigma$ value on radio maps. Under these assumptions, we have
re-determined the radio-MIR correlation through ASURV, obtaining: 
\begin{eqnarray}
~~log (L_{1.4~GHz}/L_{\odot}) = (1.08 \pm 0.04) log (L_{15~\mu m}/L_{\odot}) \\\nonumber
  - (6.07 \pm 0.41)
\label{radio_mir_lum}
\end{eqnarray}
\nin with a dispersion of $\sim$0.34 dex. This relation, shown as dot-dashed line in
the $left$ panel of figure \ref{fig_ricorr}, is somewhat more scattered than 
and lies about a factor of 2 below the previous determination.
The $right$ panel of figure \ref{fig_ricorr} shows the radio to MIR luminosity
ratio versus MIR luminosity.
Our estimate of the ``real'' correlation (dot-dashed line) has a lower normalization 
than the local radio--FIR correlation (dotted line; extrapolated to radio--MIR as described 
below), which is instead much closer to our determination for detections only (dashed line).
The local relation corresponds to a local value of the ``$q$'' parameter 
equal to 2.34 (defined as $q \equiv log (L_{FIR} [W] / (3.75 \times 10^{12} [Hz]) \times
1 / L_{1.4~GHz} [W Hz^{-1}])$, where the FIR flux is defined to be $1.26 \times 10^{-14} 
(2.58 S_{60 \mu m} + S_{100 \mu m})$ [W m$^{-2}$]; see Condon et al. 1991)
and is converted to MIR through the empirical relations between $L_{FIR}$ and 
$L_{15~\mu m}$ found for local galaxies by Elbaz et al. (2002). 
In the same figure we also show the relation derived (using the same K-corrections and
cosmological parameters considered in our analysis) from the radio and 15-$\mu$m
data for 19 ISOCAM sources detected in the WSRT deep radio survey of the HFD-N region
(see Table 1 in Garrett 2002). 
This radio--MIR correlation (shown as dot-dot-dot-dashed 
line in the figure) has a normalization
significantly higher (about a factor of 5) than our best-fitting relation and is
also higher than the local one. Note, however, that this relation is derived using only
the radio detections with spectroscopic $z$, without taking into account the radio
upper limits. Since the number of upper limits is similar to that of the detections
(which constitute $\sim$40\% of the 15-$\mu$m sample detected at $> 5 \sigma$ level), 
it is likely that, as we find for our data, the ``real'' correlation
would have a somewhat lower normalization.
Despite this, the comparison of our data with those in the HDF-N region suggests a 
possible change in normalization of the radio--MIR correlation at the higher redshifts
sampled by the HDF-N ISO selected galaxies.
Alternatively, the two sets of data (ELAIS and HDF-N surveys) could be made consistent
with each other if the ISO or the radio data in the two fields were on a different flux scale
(i.e. underestimated ISO or overestimated radio flux densities in the HDF-N). 

Despite the difference in normalization between our radio--infrared
correlation and the local one, which seems to suggest a change in the
radio--infrared correlation with redshift, the important result of our
analysis is that radio and MIR luminosities for galaxies strongly
correlate with each other, almost as well as that found for  the radio and FIR,
and at significantly higher redshifts than those explored by IRAS.
This implies a possible correlation between the PAH emission and the radio 
(and FIR) luminosity. The somewhat larger
dispersion with respect to that observed for the radio--FIR relation is
due to the large spread in the mixture of PAH and
thermal emission at the high temperatures responsible for the radiation
observed in the MIR band, or to the complicated shape of the galaxy
SEDs in that wave-band, which introduces uncertainties and significant
object-to-object variations in the MIR K-correction.

\begin{figure}
\vspace{-0.4cm}
\centerline{\psfig{figure=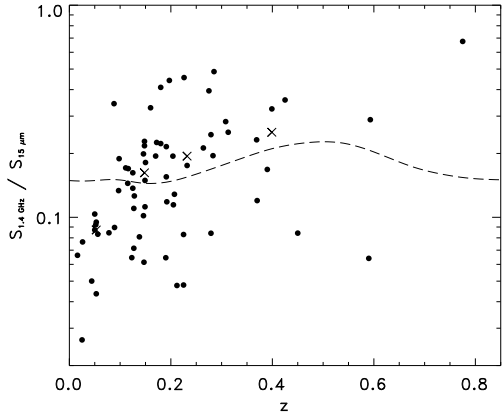,width=7.5cm}}
\caption{Radio/MIR flux density ratios versus redshift for our spectroscopically
identified galaxies. Diagonal crosses are the median values of the ratio in different 
redshift intervals ($z<0.1$, $0.1\leq z<0.2$, $0.2\leq z<0.3$ and $z\geq 0.3$), plotted 
in correspondence of the median redshift. The dashed line shows the expected change with $z$ 
of S$_{1.4~ GHz}/S_{15~\mu m}$ due to the difference between the average radio and
MIR starburst K-corrections (normalised to the median value for our data, corresponding
to S$_{1.4~ GHz}/S_{15~\mu m}=0.15$) at $z=0$.}
\vspace{-0.6cm}
\label{fig_fluxz}
\end{figure}

Figure \ref{fig_fluxz} shows the ratio between 1.4-GHz and 15-$\mu$m
flux densities as a function of redshift for our spectroscopically identified
sample of galaxies with radio detections. The dashed line shows the
expected change of this ratio due to the different K-corrections (for
starburst galaxies) in the two bands. The normalization of this curve
has been chosen so that approximately the same number of objects lie
above and below the line. This corresponds to a value $<S_{1.4 GHz} / S_{15
  \mu m}> \simeq 0.15$ at $z = 0$.  This value is in good agreement
with the value obtained by combining the well known 1.4-GHz / 60-$\mu$m
relation -- $S_{60 \mu m} \simeq 127~ S_{1.4 GHz}$ (Cram et al. 1998)
-- and the average 15-$\mu$m / 60-$\mu$m ratio found by Mazzei et al.
(2001) and Xu (2000) -- $<S_{15 \mu m} / S_{60 \mu m}> \simeq 0.05$
(note however above that our radio detections
describe the upper envelope -- i.e. stronger radio sources -- of the
radio--MIR correlation, rather than the entire population).
Despite the relatively large spread in the values of the radio--MIR
flux density ratio, no obvious trend with redshift is seen in our data for $z
\gsimeq 0.07$.  All the objects below this redshift show values of
their radio to MIR flux density ratios which are significantly lower than the
mean. One possible reason for this might be that we have missed some
extended emission in some of these sources. In fact, about
half of the very low redshift sources have a weak radio emission
(detected below the $5\sigma$ threshold) therefore, because of the low
signal-to-noise ratio, their flux density may have been underestimated if
their radio emission is more extended than the beam ($\sim 15^{\prime
  \prime}$). Indeed, most of these faint radio sources at low redshift
are bright and extended in both optical and MIR ($\sim 40^{\prime
  \prime} - 1^{\prime}$).

\begin{figure*}
\begin{minipage}{17.5cm}
\vspace{-0.2cm}
\centerline{\psfig{figure=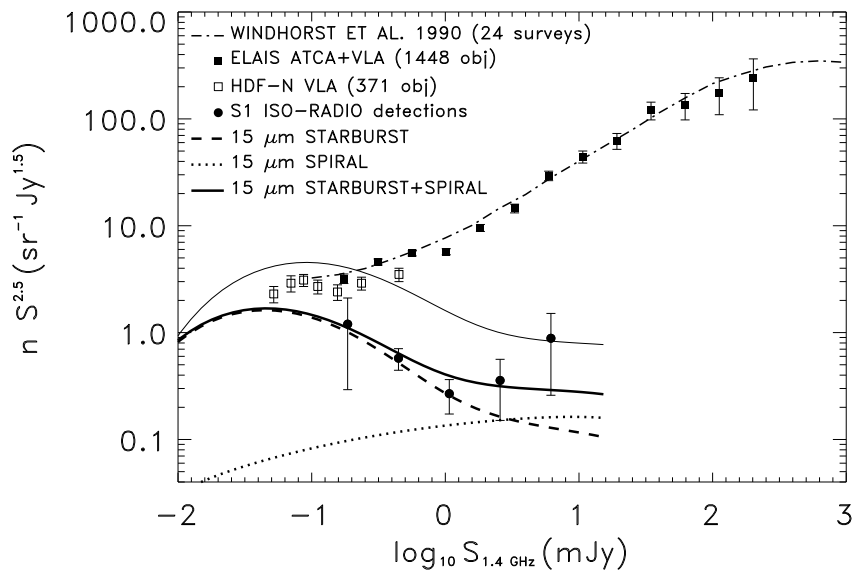,width=12cm}}
\caption{Best-fitting model to our ELAIS 15-$\mu$m extragalactic source counts (in differential 
form, normalised to a Euclidean non-evolving distribution) converted to 1.4 GHz by considering 
the empirical relation found for our data and its spread. 
As explained in the plot, the thick
solid line represents the expected total source counts, while the short-dashed and the dotted
lines are respectively the modeled contributions of a population of strongly 
evolving starburst galaxies (plus Seyfert 2) and non-evolving spirals. The thin solid line is the
total contribution from starburst and spiral galaxies obtained by increasing the radio--MIR
relation normalization by a factor of 2. 
The filled circles are the counts of the S1 15-$\mu$m sources with a $\geq 3 \sigma$ 1.4-GHz 
counterpart. The dot-dashed line represents the fit of Windhorst, Mathis \& Neuschaefer (1990) to the
1.4-GHz counts obtained from 24 different radio surveys. The filled squares are the total radio counts 
in the ELAIS regions (a combination of the S1 ATCA data of Gruppioni et al. 1999 with the 
VLA data in the northern ELAIS regions of Ciliegi et al. 1999). The open squares are the 
1.4-GHz counts in the HDF-N from Richards (2000).}
 \label{fig_rcnt}
\vspace{-0.3cm}
\end{minipage}
\end{figure*}

\section{Contribution of MIR Galaxies to the Radio Source Counts}
\label{radio_cnt}
In order to estimate the contribution made to the radio source counts 
from infrared galaxies,
we have convolved the correlation between radio and 15-$\mu$m
luminosities derived in the previous section with the evolutionary
model for spiral and starburst galaxies which fits the 15-$\mu$m source
counts in S1 (Gruppioni et al. 2002). 
The intrinsic dispersion adopted
for the radio--MIR correlation ($\sigma\sim0.28$ dex) has been derived
by subtracting from the observed value ($\sim$0.34 dex) the estimated
contributions from the uncertainties on the radio and 15-$\mu$m
observed flux densities ($\sim 0.08$ and 0.10 dex, respectively) and the radio
and 15-$\mu$m K-corrections ($\sim$0.02 and 0.15 dex, respectively).
The resulting predicted source counts for
starburst + Seyfert 2, normal spiral and all galaxies, are plotted in
figure \ref{fig_rcnt} as dashed, dotted and solid lines respectively. 
As a consistency
check, we have derived the radio counts of our ISO selected galaxies
directly from the data. Each radio detected ISO galaxy has been
weighted by its radio and MIR spatial coverage and the contribution from
all the radio detections in radio bins have been summed to produce the
counts shown as filled circles in figure \ref{fig_rcnt}. The data
points are in excellent agreement with the model predictions; in
particular, the star-forming population would be responsible for about
40\% -- 60\% of the observed counts at $S_{1.4 GHz} \sim 50 - 100~
\mu$Jy. Therefore, starburst galaxies would make up most of the
observed radio counts at $\mu$Jy level, in agreement with the results
from very deep radio surveys, like that in the HDF-N, where $\sim 80$\%
of sources detected at $S_{1.4~GHz} > 16 ~\mu$Jy have been identified
with starburst galaxies (Richards (2000)). Conversely, the
fractional contribution of starburst galaxies to the radio counts
decreases rapidly above $S_{1.4 GHz} \sim 0.1$ mJy; in good agreement
with spectroscopic identifications at the sub-mJy level, which 
find $\sim$60--70\% of elliptical galaxies and AGN1 among the optical
counterparts of $S_{1.4~GHz} \gsimeq 0.2$ mJy radio sources (Gruppioni,
Mignoli \& Zamorani 1999).

Given the differences in the normalization of the relations between
radio and MIR luminosities derived from different samples (see previous
section), we have computed the expected radio counts by increasing the
normalization of our best-fitting correlation by a factor of 2, in such
a way to bring it on the same scale as the local correlation. The
resulting (total) counts are shown by the thin solid line in figure
\ref{fig_rcnt}. The higher normalization relation produces a
contribution to the radio counts which is far too high with respect to
the observed data and exceeds the total radio counts at $S_{1.4 GHz}
\lsimeq 0.3$ mJy. Of course, this result is produced by the combination
of the radio-MIR correlation and a model fitting the source counts.
However, the model considered here is the only one able to fit the
15-$\mu$m counts in S1, which are lower than the other existing ones.
Therefore, a radio-MIR correlation with a normalization significantly
higher than that found in this work seems to be inconsistent with the
observed faint 1.4-GHz source counts.

\section{Conclusions}
\label{concl}
By searching for 3$\sigma$ detections on deep ATCA 1.4-GHz maps of the
southern ELAIS fields S1 and S2, and cross-correlating these with
15-$\mu$m extragalactic objects detected respectively by Lari et al.
(2001) and by Pozzi et al. (2002, in preparation), we have obtained a
sample of 84 MIR-radio galaxies (65 of which have measured redshifts).

These data have allowed us for the first time to study with a statistically
significant sample of objects, the radio--MIR correlation and the radio
and infrared properties of galaxies to much larger distances and
fainter flux densities than previously achieved with IRAS.

The principal results of our analysis are the following:
\begin{itemize}
\item[1.] The radio--MIR correlation for MIR selected galaxies with a
  radio detection in our radio maps is well described by an
  approximately linear relation, with a scatter of $\sim 0.27$ dex
  (similar to that found for the local radio--FIR relation), implying
  that PAH band emission correlates with FIR and radio
  luminosity and that the locally determined correlation between
  radio and infrared emission for star-forming galaxies persists
  to cosmological distances ($z \sim 0.6$).
\item[2.] If we consider also the radio upper limits we can obtain an estimate
of the ``real'' radio--MIR correlation, unbiased by the radio non-detections.
We still obtain a strong correlation between radio and MIR luminosities, but with a 
factor 2 lower normalization and a larger spread ($\sim 0.34$ dex). 
The lower normalization of our relation corrected for upper limits with respect to 
the local radio--MIR relation (derived from the radio--FIR through a local FIR/MIR 
average ratio) implies a change in the radio--MIR correlation with increasing 
redshift. 
\item[3.] There is no indication of any trend with $z$ in the radio--MIR correlation found 
for our data (apart the K-correction effects), up to $z \sim 0.6$.
\item[4.] The contribution of 15-$\mu$m galaxies to the radio source
  counts has been computed directly from our data and also by including
  the empirical radio--MIR correlation and its spread into the model
  fitting the MIR extragalactic source counts. Data and model agree
  very well and predict that MIR starburst galaxies should start
  contributing significantly ($\gsimeq$10\%) to the radio counts around
  0.5--0.8 mJy, with their importance rapidly increasing until they
  make up $> 40 - 60$\% of the observed counts at $S_{1.4} \lsimeq 50 -
  100~\mu$Jy.
\end{itemize}

\vspace{-0.4cm}
\section*{Acknowledgments}
Part of this work was supported by MIUR (Cofin) and ASI research grants.

\end{document}